\documentclass[aps,twocolumn,nofootinbib,superscriptaddress]{revtex4-2} 
\usepackage[utf8]{inputenc}
\usepackage{graphicx}
\usepackage{braket}
\usepackage{dcolumn} 
\usepackage{bm} 
\usepackage{amsmath}
\usepackage{amssymb}
\usepackage{float}
\usepackage{physics}
\usepackage{hyphenat}
\usepackage{cancel}
\usepackage{thmtools}
\usepackage{bbding}
\usepackage{xcolor}
\usepackage{soul}
\usepackage[pdftex]{hyperref}
\hypersetup{colorlinks,linkcolor={blue},citecolor={blue},urlcolor={blue}}
\usepackage{subfigure}

\begin{document}
	\title{Nonlinear three-state quantum walks}
	\author{P. R. N. Falc\~ao}
	\email{pedro.falcao@fis.ufal.br}
	\affiliation{%
		Instituto de F\'{i}sica, Universidade Federal de Alagoas, 57072-900 Macei\'{o}, AL, Brazil}
	\author{J. P. Mendonça}
	\affiliation{%
		Instituto de F\'{i}sica, Universidade Federal de Alagoas, 57072-900 Macei\'{o}, AL, Brazil}
	\affiliation{%
		Faculty of Physics, University of Warsaw, Pasteura 5, 02-093 Warsaw, Poland 
	}	
	\author{A. R. C. Buarque}
	\affiliation{%
		Instituto de F\'{i}sica, Universidade Federal de Alagoas, 57072-900 Macei\'{o}, AL, Brazil}
	\affiliation{
		Secretaria de Estado da Educação de Alagoas, 57055-055, Macei\'o, Alagoas, Brazil
	}
	\author{W. S. Dias}
	\affiliation{%
		Instituto de F\'{i}sica, Universidade Federal de Alagoas, 57072-900 Macei\'{o}, AL, Brazil}
	\author{G. M. A. Almeida}
	\affiliation{%
		Instituto de F\'{i}sica, Universidade Federal de Alagoas, 57072-900 Macei\'{o}, AL, Brazil}
	\author{M. L. Lyra}
	\affiliation{%
		Instituto de F\'{i}sica, Universidade Federal de Alagoas, 57072-900 Macei\'{o}, AL, Brazil}
	\begin{abstract}
		
		The dynamics of a three-state quantum walk with amplitude-dependent phase shifts is investigated. 
		We  consider two representative inputs whose linear evolution is known to display
		either full dispersion of the wave packet or intrinsic localization on the initial position.
		The nonlinear counterpart presents much more involved dynamics featuring
		self-trapping, solitonic pulses, radiation, and chaotic-like behavior. We show that nonlinearity leads to a metastable self-trapped wavepacket component that radiates in the long-time regime with the survival probability $\propto t^{-1/2}$. A sudden dynamical transition from such metastable state to the point when the radiation process is triggered is found for a set of parameters. 
		
	\end{abstract}
	\maketitle
	
	\section{Introduction} 
	The crucial role of stochastic processes in physics and computer science, added to the tremendous progress in the field of quantum computation in recent years, have fueled studies of quantum analogues of random walks, namely quantum walks \cite{aharanov1993}.
	Their main advantage comes from quantum coherence, allowing for interference effects that result in ballistic spreading of the walker, in contrast to the classical diffusive spreading \cite{knight2003}.
	Over the past few decades, quantum walks have been proved a valuable tool for designing quantum search algorithms \cite{shenvi2003,childs2004,childs2014,wong2015,wong2018}, carrying out quantum communication protocols \cite{zhan2014,stefanak2016,kurzynsky2011}, and realizing universal quantum computation \cite{childs2009,lovett2010}. 
	They have also been particularly useful to simulate various physical phenomena such as quantum phase transitions \cite{chandrashekar2008,wang2018,wang2019}, Anderson localization \cite{mendes2019,buarque2019,obuse2011,queiros2020}, rogue waves \cite{buarque2022}, and nonlinear dynamics \cite{shikano2014,mendonca2020,navarrete2007,buarque2020,buarque2021}. 
	
	Nonlinear models of discrete-time quantum walks have attracted a great deal of attention for much complex dynamics can turn up out of simple rules, these being
	pre-established iterated series of quantum gates. 
	Navarrete \textit{et al.} \cite{navarrete2007} proposed a nonlinear version of the optical Galton board in which a Kerr-type self-phase gain was applied at each step, giving rise to hallmarks of nonlinear behavior such as soliton collisions and chaos. Years later the same model was found to display self-trapping for specific angles of the coin operator \cite{buarque2020}.
	Other recent studies have been focusing on issues such as noise 
	tolerance \cite{buarque2021} and nonunitary transformations \cite{mendonca2020}. 
	Another motivation to study quantum walks of that kind is that
	nonlinear effects are inevitably at play in many of the platforms designed for their implementation, like photonic devices \cite{karski2009,schreiber2011,xu2018,zhan2017}, Bose-Einstein condensates \cite{xie2020}, and trapped-ion systems \cite{tamura2020}. 
	
	Therefore, it is paramount to 
	develop on ways to handle nonlinearity in discrete-time quantum walks, even more so
	in models with additional coin degrees of freedom, not yet examined in previous works. 
	While it has been a standard practice to consider a two-dimensional coin space, Inui \textit{et al.} \cite{inui2005} put forward a three-state version of the Hadamard walk in which
	the walker was allowed to stay put besides going left or right, according to its chirality.
	This brings about an eigenvalue degeneracy in Fourier space \cite{inui2005, falkner2014} that results in an intrinsic form of localization around the 
	walker's starting position. A number of studies
	followed the lead to cover limit theorems \cite{machida2015,stefanak2014}, quantum-to-classical transition \cite{tude2021}, circuit implementation \cite{saha2021}, and universal dynamical scaling laws \cite{falcao2021}. A similar class of quantum walks in which the walker is allowed to perform self-loops in each vertex of a graph was showed to improve search algorithms \cite{wong2015,wong2018}. 
	
	Here we set out to explore nonlinear mechanisms in three-state quantum walks. We are primarily interested in seeing how nonlinearity acts upon the intrinsic trapping mechanism mentioned above \cite{inui2005}. 
	We find that the localized component displays a metastable character, escaping from its initial position after a transient time through a radiation process, which is discussed in detail. 
	
	\section{Model}       
	
	We consider a discrete-time quantum walk on the line that runs the Hilbert space $H = H_p \otimes H_c$, made up by position states $\left\lbrace{|n\rangle} \right\rbrace$ in $H_p$ and 
	a three-dimensional coin space $H_c$ spanned by $\lbrace\ket{L},\ket{S},\ket{R}\rbrace$.
	The evolution of the walker's state vector is performed via
	successive applications of a unitary operator $\hat{U}$ following $\ket{\psi(t)} = \hat{U}(t)\ket{\psi(t-1)}$.
	Its standard form reads $\hat{U}=\hat{S}[\hat{C}\otimes\mathbb{I}_{P}]$, where $\mathbb{I}_p$ is the identity operator acting on the $n$-dimensional position space, $\hat{C}$ is the coin operator responsible for generating superpositions, and $\hat{S}$ is the conditional shift operator which moves the walker (wavefunction amplitudes) through the lattice according to its internal degrees of freedom. In three-state quantum walks, the walker can 
	move to the left, right, or stay at its current position, as embedded in
	\begin{eqnarray}
		\hat{S} = \sum_{n =-\infty}^{\infty}&&\bigg[ |n-1\rangle \langle n| \otimes |L \rangle \langle L | + |n\rangle \langle n| \otimes |S \rangle \langle S| \nonumber\\ 
		&& + |n+1\rangle \langle n| \otimes |R \rangle \langle R \mid \bigg].
		\label{Eq.2}
	\end{eqnarray}
	The coin we consider here is a $U(3)$ operator responsible to shuffle the internal degrees of freedom of the walker. We consider the so-called Grover coin defined by \cite{inui2005}
	\begin{equation}
		\hat{C} = \frac{1}{3} 
		\begin{pmatrix}  
			-1 & 2 & 2 \\ 
			2 & -1 & 2 \\
			2 & 2 & -1
		\end{pmatrix}.
		\label{Eq.3}
	\end{equation}
	
	In order to introduce nonlinear effects into the quantum walk model we place an amplitude-dependent phase shift during evolution \cite{navarrete2007,buarque2020,buarque2021},    
	\begin{equation}
		\hat{U}_{nl}(t) = \sum_{c} \sum_{n =-\infty}^{\infty} e^{iG(n,c,t)} \ket{n,c}\bra{n,c},
		\label{Eq.4}
	\end{equation}
	with $c = L,S,R$ and $G(n,c,t)= 2\pi\chi|\psi_{n,c}(t)|^2$, where $\chi$ is the nonlinearity strength and $\psi_{n,c}(t) = \langle n,c |\psi (t) \rangle$. Therefore, the full unitary operator 
	is rewritten as $\hat{U}(t) = \hat{S}[\hat{C}\otimes\mathbb{I}_{P}]\hat{U}_{nl}(t-1)$. The first operator of the sequence then uses information about the local wavefunction amplitudes at a 
	given instant to feedback
	it in the following round. 
	
	\section{Results}
	
	Let us first recall that the dynamics of linear three-state quantum walks is better conceived when the initial state is written in terms of the eigenvectors of $\hat{C}$ \cite{stefanak2014}:
	\begin{align}
		\ket{\sigma^+} &= \frac{1}{\sqrt{3}}\ket{L} + \frac{1}{\sqrt{3}}\ket{S} + \frac{1}{\sqrt{3}}\ket{R},\\
		\ket{\sigma_1^-} &= \frac{1}{\sqrt{6}}\ket{L} - \frac{2}{\sqrt{6}} \ket{S} + \frac{1}{\sqrt{6}}\ket{R},\\
		\ket{\sigma_2^-} &= \frac{1}{\sqrt{2}}\ket{L} - \frac{1}{\sqrt{2}}\ket{R},
	\end{align}
	where $\hat{C}\ket{\sigma^+} = \ket{\sigma^+}$ and $\hat{C}\ket{\sigma_{i}^-} = - \ket{\sigma_{i}^-}$, with $i=1,2$.  
	The dynamics obtained from state $\ket{\sigma_1^{-}}$, 
	resembles that of the standard 1D Hadamard walk featuring full dispersion
	of the wavefunction with its characteristic peaks 
	at the front pulse. 
	In contrast, $\ket{\sigma^+}$ leads to a kind of intrinsic localization of the wavefunction as we will see shortly. 
	
	\begin{figure}[t!]
		\centering
		\includegraphics[width = 0.5\textwidth]{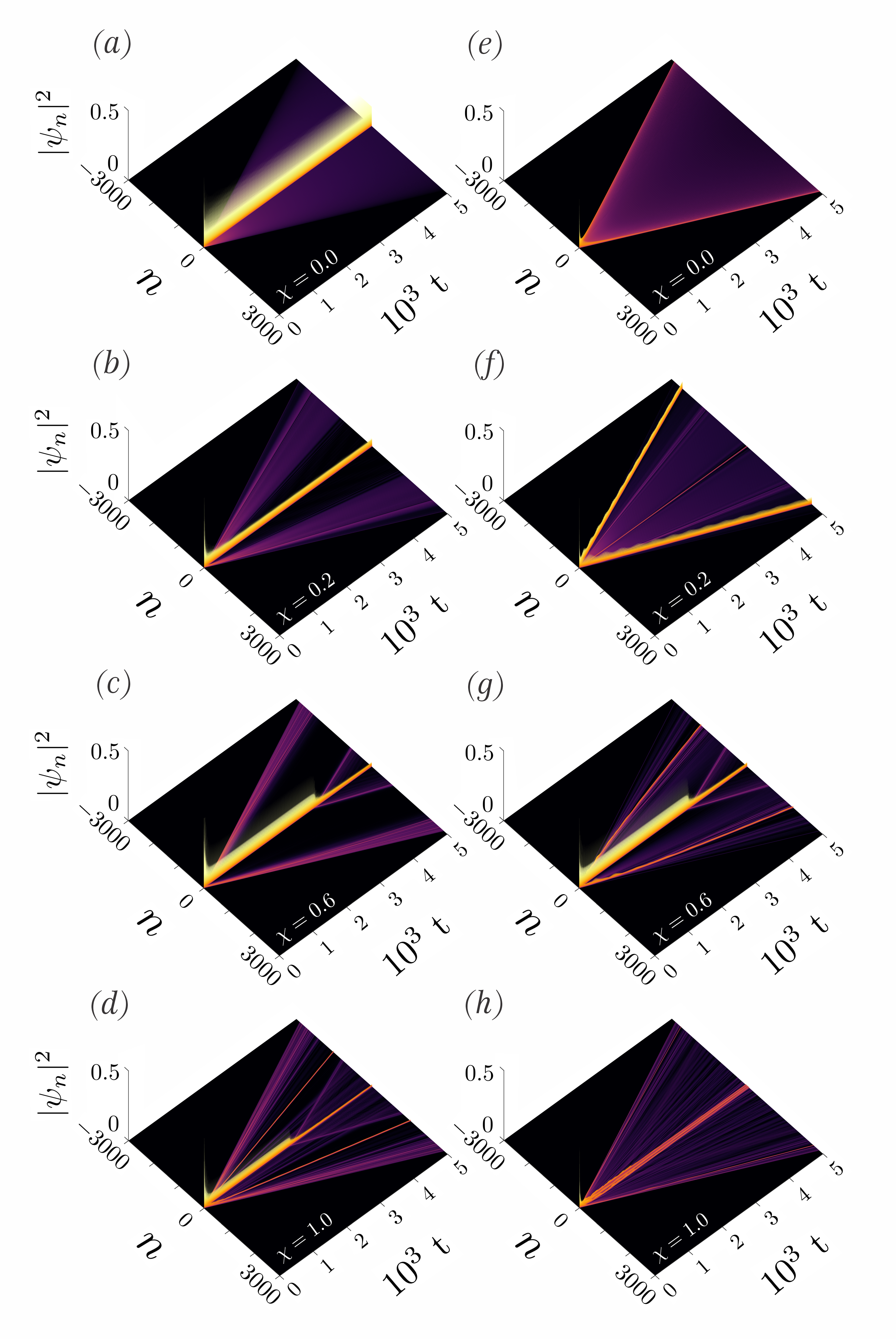}
		\caption{Time-evolution of the probability density $|\psi_n|^2$ in the position space for some representative values of the nonlinear parameter $\chi$. The coin component $\ket{c}$ of the input is set as
			(a-d) $\ket{\sigma^+}$ and (e-h) $\ket{\sigma_1^{-}}$.  From top to bottom $\chi = 0, 0.2, 0.6, 1.0$. Note that nonlinearity brings the localized component to a metastable state and induces some degree of localization when $\ket{c}=\ket{\sigma_1^{-}}$.}		
		\label{Fig.1}
	\end{figure}
	
	We shall now focus our attention on the influence of nonlinear phase shifts on the dynamics by
	initializing the state of the walker as $\ket{\psi(0)} = \ket{n=0}\otimes \ket{c}$, with $\ket{c}$ set in either $\ket{\sigma^+}$ or $\ket{\sigma_1^{-}}$. Figure \ref{Fig.1} shows the resulting evolution of the probability density $|\psi_n|^2 = |\psi_{n,L}|^2+|\psi_{n,S}|^2+|\psi_{n,R}|^2$ for different values of $\chi$.
	When only linear effects are taken into account ($\chi = 0$), the walker becomes strongly localized around the starting position for $\ket{c} = \ket{\sigma^+}$, as expected [Fig. \ref{Fig.1}(a)]. For $\ket{c} = \ket{\sigma_1^{-}}$, the walker basically mimicks the two-state Hadamard walk [Fig. \ref{Fig.1}(e)].
	The situation changes drastically when the nonlinear contribution sets in. For weak nonlinearity, say $\chi = 0.2$, and $\ket{c}=\ket{\sigma^+}$, the share of the wavefunction amplitude surrounding the origin is much lower than in the linear case as a couple of outgoing pulses build up. 
	Much pronounced soliton-like structures are seen when $\ket{c}=\ket{\sigma_1^{-}}$ as 
	a self-trapped component starts to develop.
	As $\chi$ is increased,
	the localized component is found to remain stable during some transient time until
	it starts to radiate. This behavior holds for both coin inputs [see Figs. \ref{Fig.1}(c,g)].
	Even more complex patterns of soliton formation, wave-packet radiation, and self-trapping take place as we further increase the strength of nonlinearity [see Figs. \ref{Fig.1}(d,h)].
	
	\begin{figure}[t!]
		\centering
		\includegraphics[width = 0.49\textwidth]{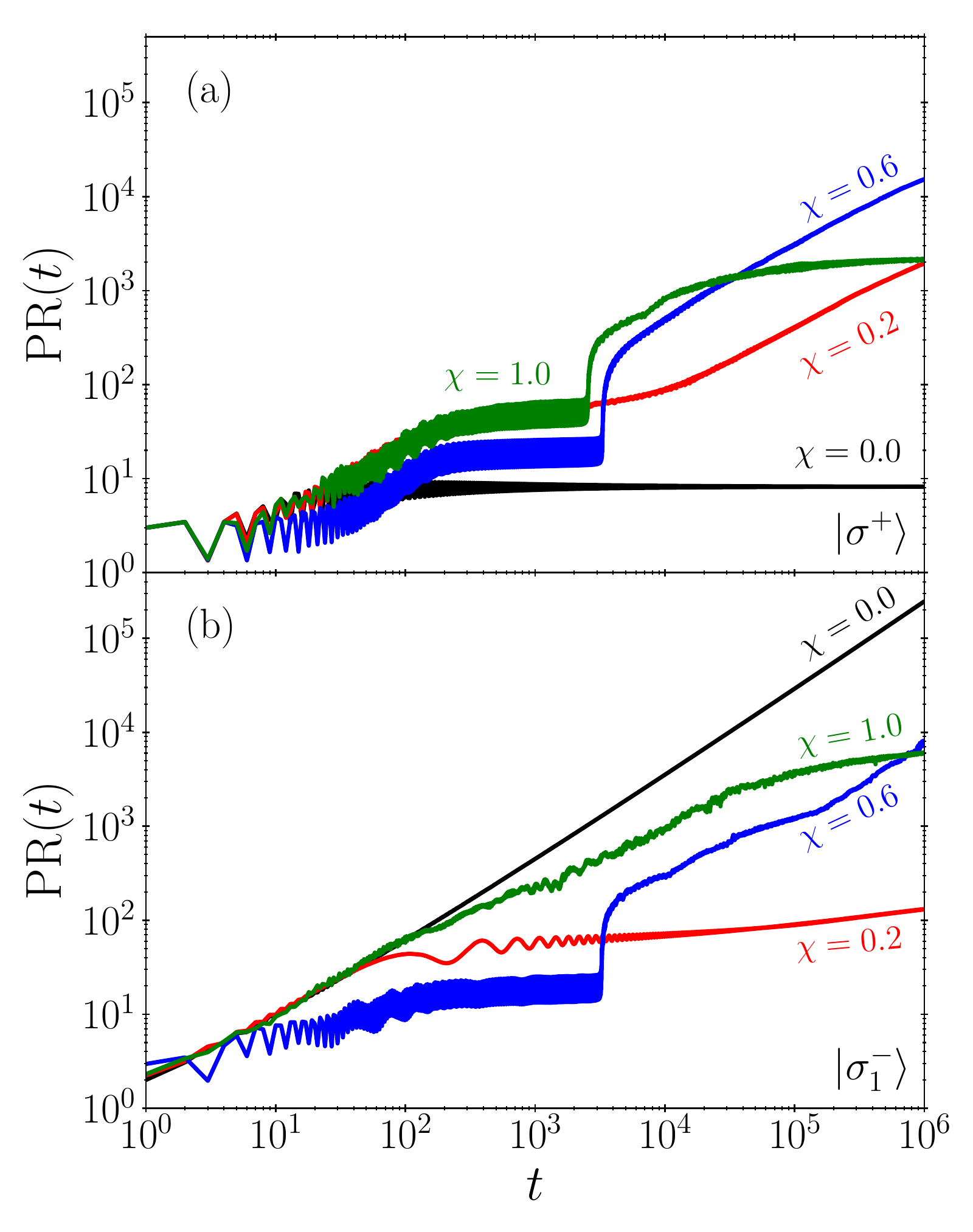}	
		\caption{Time evolution of the participation ratio for the coin input states (a) $\ket{\sigma^+}$ and (b) $\ket{\sigma_1^-}$. This quantity offers a clear signature of the radiation dynamics when nonlinear effects are taken into account.}	
		\label{Fig.2}		
	\end{figure}
	
	For a better characterization of the radiation phenomenon, 
	it is useful to take a look at the participation ratio 
	$\text{PR}$, which accounts for how sparse the wavefunction is:
	\begin{equation}
		\text{PR}(t) = \frac{1}{\sum_{n} \mid \psi_n(t) \mid ^{4}},
		\label{Eq.7}
	\end{equation}
	where the sum runs over all lattices sites. Its evolution is depicted in Fig. \ref{Fig.2} for the same couple of coin inputs and values of $\chi$ as before. 
	In the absence of nonlinearity, considering the coin input that leads to localization, the participation ratio saturates to a level about the width of the trapped component. 
	In the nonlinear case, we spot two distinct regimes. The first one lasts for a short term during which the participation ratio remains roughly constant, embodying a signature of localization. In the long-time regime, $\text{PR}$ grows continuously, indicating spreading of the wavefunction. The detrapping of the localized component can occur suddenly (for strong nonlinearities) or smoothly (typically for weak nonlinearites). 
	It is worth stressing that series of detrapping dynamic transitions may develop for strong nonlinearities. Furthermore, the formation of solitonic structures or localized chaotic-like pulses also leads to the saturation of $\text{PR}$ \cite{navarrete2007,buarque2020}. 
	When $\ket{c}=\ket{\sigma_1^{-}}$, for $\chi=0$,
	the evolution of $\text{PR}$ is linear in time with a logarithmic correction, as recently reported in \cite{falcao2021}. In the presence of nonlinearity, 
	again, $\text{PR}$ does not change much in the short term and this is due to two distinct aspects. While for low values of $\chi$ it tells the formation of outgoing solitonic structures at the wave front [cf. \ref{Fig.1}(f)], for strong $\chi$ it is 
	mostly due to the trapped component [cf. \ref{Fig.1}(g,h)].
	
	\begin{figure}[t!]
		\centering
		\includegraphics[width = 0.49\textwidth]{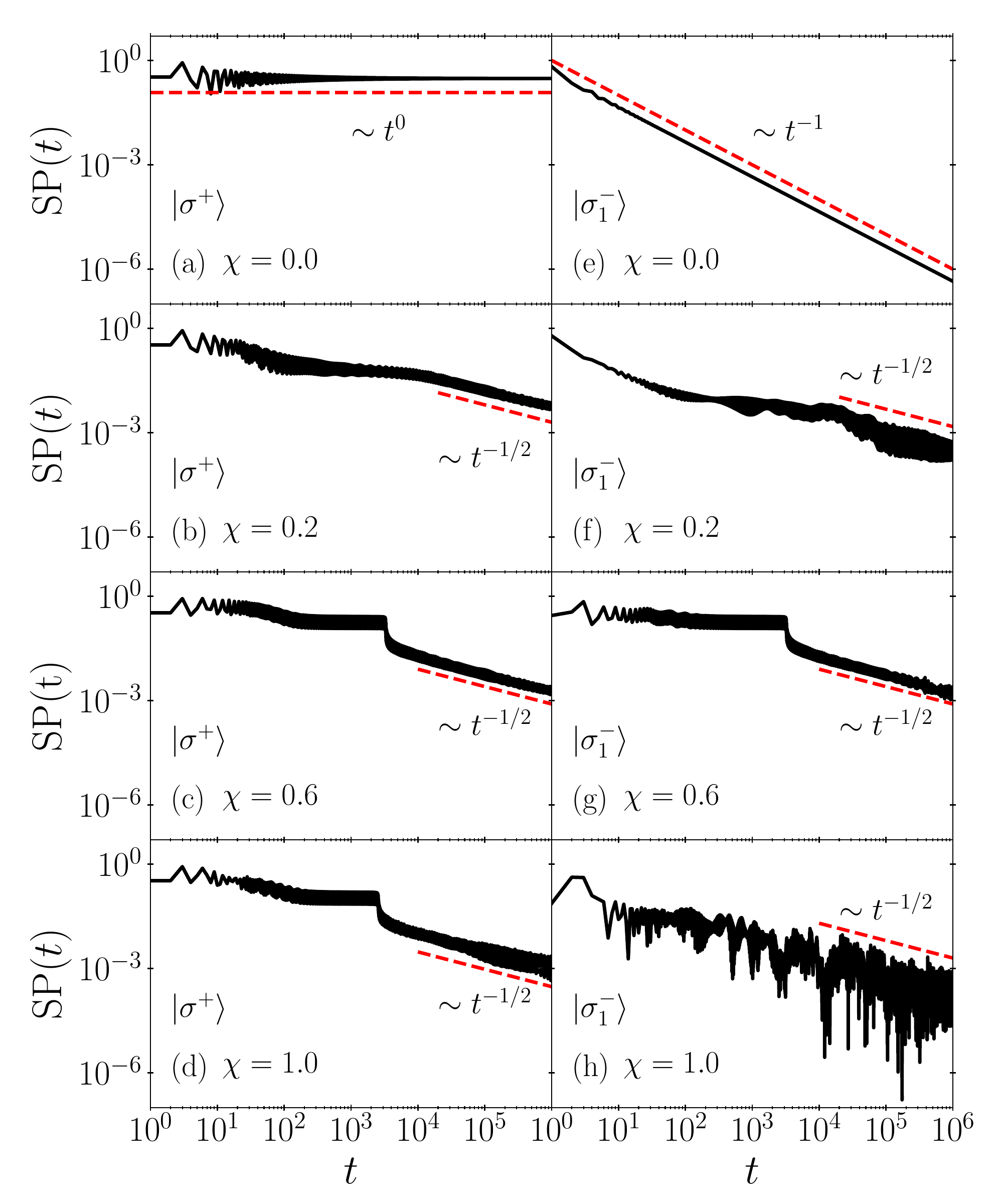}
		\caption{Survival probability versus time steps for the usual coin inputs, $\ket{\sigma^+}$ and $\ket{\sigma_{1}^{-}}$, and different values of $\chi$. We note that radiation dynamics of the localized pulse approximately follows a power law.}		
		\label{Fig.3}		
	\end{figure}
	
	\begin{figure}[t!]
		\centering
		\includegraphics[width = 0.49\textwidth]{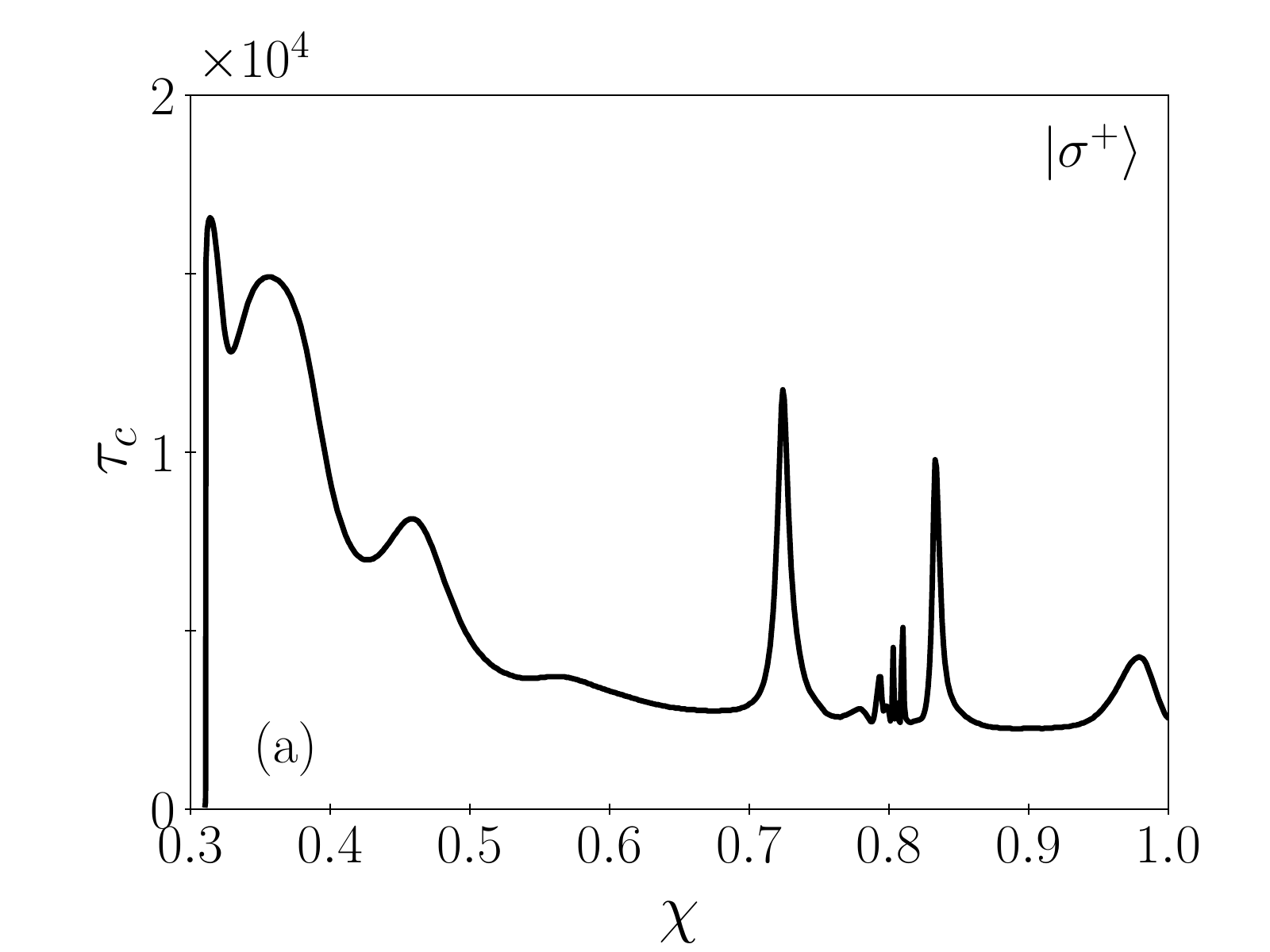}
		\includegraphics[width = 0.49\textwidth]{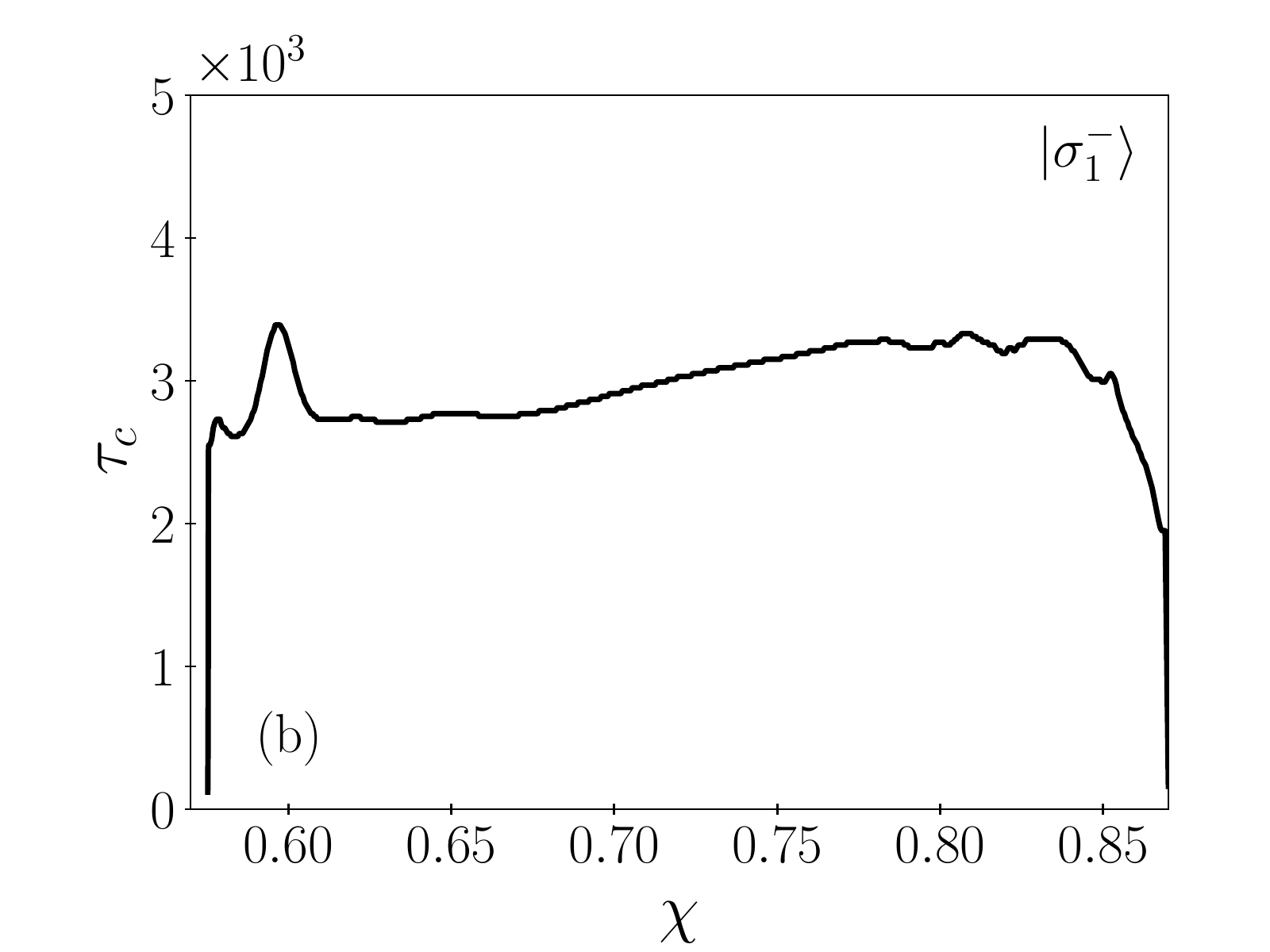}
		
		\caption{Duration of the metastable state $\tau_c$ as a function of the nonlinear strength $\chi$ for both coin inputs. Note the pronounced peaks for specific values of $\chi$ when $\ket{c}=\ket{\sigma^+}$.}			
		\label{Fig.4}		
	\end{figure}
	
	Now, in order to analyze the radiation dynamics from a local perspective, we compute the evolution of the survival probability $\text{SP(t)}=|\psi_{n=0}(t)|^2$, which is displayed in Figs. \ref{Fig.3}(a-d) and \ref{Fig.3}(e-h) for the coin inputs $\ket{\sigma^+}$ and $\ket{\sigma_{1}^{-}}$, respectively. When the quantum walk is purely linear, the first renders saturation of the $\text{SP(t)}$, whose level can be easily computed analytically as demonstrated in \cite{stefanak2014}. For the other input, we find $\text{SP} \propto t^{-1}$, as expected. 
	Their nonlinear counterpart is mainly characterized by a short-time regime during which the survivor probability oscillates around a constant value and a long-time regime featuring in which $\text{SP}$ decays slowly due to the radiation process. Therein, we find $\text{SP} \propto t^{-1/2}$, although this trend
	is obscured with random fluctuations for extreme values of $\chi$.

	
	We have seen that  
	the triggering of the radiation process can occur abruptly for specific sets of parameters. For these cases, we evaluate 
	how long, $\tau_c$, the trapped component is able to hold on to its metastable state. To do that, we track the evolution of the participation ratio up to the moment its slope suddenly increases (see Fig. \ref{Fig.2}). The result is plotted in Fig. \ref{Fig.4} against $\chi$, set within the appropriate ranges.
	Overall, when the initial state features $\ket{c}=\ket{\sigma^+}$, the metastable state lives longer 
	and does so over a wider range of nonlinearity strengths. There are some particular values of $\chi$ for which the self-trapped component is significantly more stable. This is related to the complex structure of the dynamical attractor, typically observed in nonlinear dynamical systems presenting a chaotic-like dynamics. We discuss it right next.  
	
	\begin{figure}[t!]
		\centering
		\includegraphics[width = 0.49\textwidth]{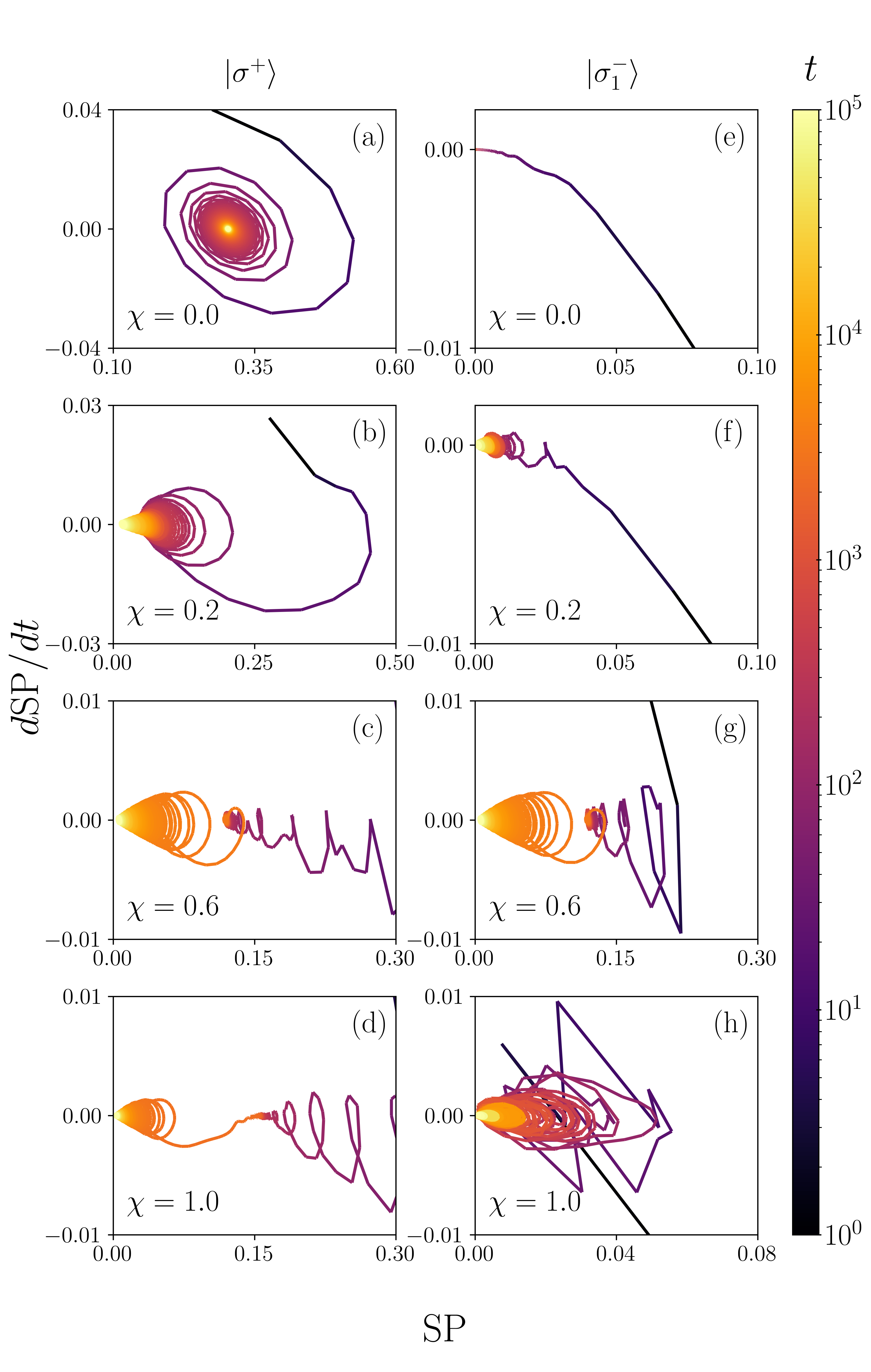}
		\caption{Phase portraits $\text{SP} \times d \text{SP}/dt$ for the coin inputs $\ket{\sigma^+}$ (left column) and $\ket{\sigma_1^{-}}$ (right column), and distinct values of $\chi$. The color gradient is to give a sense of direction. Note that the localized component becomes unstable in the presence of nonlinearity. Metastable self-trapping and chaotic-like dynamics comes about in some scenarios.}			
		\label{Fig.5}		
	\end{figure}
	
	Finally, we offer a global view of the dynamical regimes addressed above 
	by working out some phase portraits of the survival probability $\text{SP}$ against its ''speed``
	$d \text{SP}/dt$, as shown in Fig. \ref{Fig.5}. 
	In the linear three-state quantum walk ($\chi=0$) the distinction between the portraits of the localization dynamics originated from $\ket{\sigma^+}$ and the dispersive one we get from $\ket{\sigma_1^{-}}$ is striking. 
	Now, in the presence of nonlinearity, the orbit always converges asymptotically to zero survival probability, whatever its strength, albeit carrying some differences.
	A small degree of nonlinearity ($\chi=0.2$) destabilizes the self-trapped state and full release from the origin is eventually achieved in the long run.
	However, for intermediate values of the nonlinear strength ($\chi=0.6$) one clearly observes the emergence of metastable cycles around a self-trapped state prior its release.  At last, for strong nonlinearity ($\chi=1$), the overall dynamics looks the same when $\ket{c}=\ket{\sigma^+}$ whereas
	a chaotic-like trajectory from the start to full release sets in when $\ket{c}=\ket{\sigma_1^{-}}$.

	\section{Concluding Remarks}
	
	We studied the role of a amplitude-dependent phase modulation on the dynamics of a three-state quantum walk.
	Such nonlinear contribution was shown
	to provide with much involved dynamics, with distinct paths depending on the input. %
	
	By tracking down the evolution of the walker over the range of nonlinearity strengths, we showed that
	the self-trapped state that readily develops at the initial position becomes unstable, with its survival (return) probability amplitude decaying asymptotically in time as $\text{SP}\propto t^{-1/2}$ as it radiates.
	We found the transition from such metastable state to the point it starts to radiate can occur abruptly for a range of $\chi$ values.
	The metastable regime could be properly identified for a wide (narrow) range of nonlinear strengths for the coin input  $\ket{\sigma^+}$ ($\ket{\sigma_1^{-}}$) and found to last for about $10^3-10^4$ time steps.   
	
	While we have mostly focused on
	the interplay between the intrinsic localization (or lack thereof) that occurs 
	in the linear regime and the onset of self-trapping with its radiation process, our work has a greater appeal. 
	Just like simple one-dimensional maps often encountered in the field of nonlinear systems give rise to incredibly complex dynamics, discrete-time quantum walks are a versatile way to simulate a wide class of phenomena. The freedom involved in setting up the unitary operator (or even nonunitary if you will \cite{mendonca2020}), input, nonlinearity profile, and internal degrees of the walker makes it a compelling case. 

	\section{Acknowledgments}
	This work was supported by CAPES, CNPq, and FAPEAL (Alagoas State research agency).

\end{document}